\documentclass[twocolumn,tighten,time]{aastex701}
\usepackage{hyperref}
\usepackage[fleqn]{amsmath}
\usepackage{graphicx}	
\pdfoutput=1 
\usepackage{epsfig}
\usepackage{subfigure}
\usepackage{multirow}

\newcommand{\der}{{\rm d}} \newcommand{\ph}{_{\rm p}}
 
\newcommand{\x}{^{\rm x}}  
 \newcommand{\s}{_{\rm s}} 
 \newcommand{\m}{_{\rm m}}  
\newcommand{\R}{{R_{\rm f}}}  
 \newcommand{\cc}{_{\rm c}}

 \newcommand{\T}{{\Re}} 
 \newcommand{\h}{_{\rm h}}

\newcommand{\F}{^{\rm th}}  
\newcommand{\G}{{\cal G}} 
\newcommand{\p}{_{\rm p}} 
\newcommand{\g}{_{\rm g}} \newcommand{\lo}{_{\rm l}}

\newcommand{\modot}{M$_\odot$\ } \newcommand{\modotc}{M$_\odot$}
\newcommand{\ti}{t_{\rm i}} \newcommand{\e}{^{\rm e}}
\newcommand{\ii}{_i}

\newcommand{\beq}{\begin{equation}} \newcommand{\eeq}{\end{equation}}
 \newcommand{\ff}{_{\rm c}}
 \newcommand{\beqa}{\begin{eqnarray}}
\newcommand{\eeqa}{\end{eqnarray}} \newcommand{\lav}{\langle}
\newcommand{\rav}{\rangle} 
\newcommand{\vir}{_{\rm vir}} 
 
 \newcommand{\f}{_{\rm f}}

  \newcommand{\Lag}{^{\rm L}}

\begin{document}

\shorttitle{Rotational Halo Properties}
\shortauthors{Salvador-Sol\'e \& Manrique}

\title{THE TIDAL TORQUE THEORY REVISITED. II. ROTATIONAL HALO PROPERTIES}

\author{Eduard Salvador-Sol\'e and Alberto Manrique}
\affiliation{Institut de Ci\`encies del Cosmos. Universitat de Barcelona, E-08028 Barcelona, Spain}

\email{e.salvador@ub.edu}

\begin{abstract}
The peak model of structure formation was built more than fifty years ago with the aim to address the origin of dark matter halo rotation in the tidal torque theory (TTT). Paradoxically, it has allowed one to explain and reproduce all halo properties found in cosmological simulations except their rotation, which remains to be understood. With the present two Papers we remedy this anomaly. In Paper I we derived the angular momentum (AM) of protohalos centered on triaxial peaks of suited scale, taking into account that, to leading order, their density profile is smooth and homogeneous. Here we use that result to derive the AM of these objects, accounting for the fact that their actual density profile is slightly outward decreasing and lumpy so that they do not collapse monolithically at once, but progressively from inside out, undergoing mergers during the process. By monitoring in detail their resulting mass and AM growth, we characterize the spin distribution of final halos and the precise mass and radial distribution of their inner mean specific AM. The results obtained explain and reproduce the rotational properties of simulated halos. 
\end{abstract}

\keywords{methods: analytic --- cosmology: theory, dark matter --- dark matter: halos --- galaxies: halos}


\section{INTRODUCTION}\label{intro}

Cold dark matter (CDM) halos play a crucial role in galaxy formation. Given the difficulty of dealing analytically with hierarchical clustering, their properties have been determined by means of cosmological simulations (see the review by \citealt{FW12}). But understanding how they are set requires analytic modeling.

Some authors tried to find their origin in smooth monolithic collapse or pure accretion (\citealt{GG72,AR98,NS99,dPea00,Bea01,Mea03,Aea04,Sea07}). Others focused on the effect of cumulative mergers (\citealt{SW98,Sea00a,G01,Mea02,Dea03,Vea02,Bea12a,Bea16}). And still others \citep{ssm,Rea98} advocated for a hybrid scenario.  

Among all halo properties, those related with their rotation, namely: i) the dependence of their angular momentum (AM) on mass (e.g. \citealt{EJ79,BE87,Cea96b,Sea00,Liao17}), ii) the nearly universal lognormal spin distribution \citep{BE87,CL96,VdB02,SS05,Aea05,Bea07,GY07,Mea07,Zs17}, and iii) the inner specific AM distribution \citep{BE87,Bea01,Bea10,Liao17}, are particularly challenging to understand since, added to the general clustering issue, there is the unknown origin of halo rotation itself. 

Several mechanisms have been proposed for it (\citealt{Lea12,Cea15,Lea15,Nea20}), but the most natural one is the tidal torque of neighboring mass fluctuations on {\it aspherical} protohalos, the so-called tidal torque theory (TTT; \citealt{P69,D70,W84}). Unfortunately, the extended Press-Schechter (EPS) model of structure formation \citep{BCEK,LC} did not account for the shape of protohalos, so the need to check the validity of TTT motivated the build up of the new peak model \citep{D70,BBKS}, where protohalos are density maxima (peaks) in the random Gaussian density field smoothed with a Gaussian window of suited scale, and, as a consequence, they are triaxial. 

By implementing TTT in the peak model, \citet{H88}, \citet{HP88}, and \citet{Cea96a} calculated the typical protohalo AM. But, apart from some technical as well as practical problems (see \citealt{SM25}, hereafter Paper I), their treatment only held in the linear regime, so it could not deal with the late collapse phase where protohalos undergo shell-crossing and major mergers. The only attempt in this direction was made by \citet{Bea01}, \citet{Mea02}, and \citet{MD02}, who explored the possibility that the rotational halo properties resulted from consecutive minor mergers. 

An important step towards the accurate analytic derivation and understanding of all halo properties including their rotational ones was taken by \citet{Gea07}, hereafter GSMH, who showed that, provided halos grew inside-out during accretion with no apparent imprint of major mergers, as claimed by \citet{ssm} and \citet{Rea98}, adopting the mass accretion rate predicted in the EPS model together with the AM growth rate implied by the observed constant halo spin automatically led to the halo density, kinematic, and specific AM profiles found in simulations. 

Despite such a remarkable result, that work was rejected for publication, arguing that the inside-out growth of accreting halos and the neglect of the effects of major mergers were unjustified. Yet, at that moment there was already growing evidence, reinforced in the years to come, that accreting halos do develop inside-out \citep{FM01,Wea02,LP03,Zetal03,Sea05,Luea06,RD06,DKM07,Cea08,Wea11,LEtal13}, and that major mergers do go unnoticed in halo properties \citep{Mea99,Hea99,Mea03,Haea06,WW09,Bea12}.

Both conditions were finally proven by \citet{Sea12a} and \citet{SM19}, respectively, using the {\it ConflUent System of Peak trajectories} (CUSP) formalism \citep{MSS95,MSS96,Mea98}. That opened the possibility to derive analytically and explain, in thee peak model, the density profile (\citealt{Sea12a,Vea12,Sea23}), kinematic profiles and shape (\citealt{Sea12b}), substructure \citep{I,II,III,IV}, mass function \citep{Jea14b}, and primary and secondary biases \citep{SM24,Sea24} of simulated halos, directly from peak statistics, with no free parameters.

But to derive the rotational properties one should first implement TTT in the peak model in a rigorous practical fashion. That was done in Paper I. Here, taking up the GSMH's procedure, we derive for the first time the detailed rotational halo properties, accounting accurately for the non-linear evolution of protohalos. 

The layout of the Paper is as follows. In Section \ref{CUSP} we remind the CUSP formalism and the derivation of the mass growth of accreting halos. In Section \ref{linear} we infer their AM growth. These results are used in Sections \ref{AM} and \ref{specific} to derive the typical halo spin and their inner mean specific AM distribution, respectively. A summary of the work is given in Section \ref{sum}.

\section{Mass Growth of Accreting Halos}\label{CUSP}


As discussed in \citet{SM19}, peaks in the linear Gaussian-smoothed density field at any arbitrary initial time $\ti$ are triaxial, so protohalos undergo ellipsoidal collapse. This means that their collapse time $t$ depends not only on their density contrast $\delta$, like in top-hat spherical collapse, but also on the scale $R$, curvature $x$, ellipticity $e$, and prolateness $p$ of the peak. However, $e$ and $p$ depend only on $x$, whose probability distribution function (PDF) for peaks with fixed $\delta$ and $R$ is sharply peaked \citep{BBKS}, so all protohalos with fixed $\delta$ and $R$ collapse nearly at the same time $t$. Thus, neglecting the small scatter in the collapse times, for any given $\delta(t,\ti)$ relation, we can find the radius $R(M,t,\ti)$ of the Gaussian window such that protohalos associated with peaks at $\ti$ with those $\delta$ at $R$ give rise to halos with mass $M$ at $t$. In other words, there is essentially a one-to-one correspondence between halos with $M$ at $t$ and peaks with $\delta$ at $R$, like in top-hat spherical collapse. Obviously, neglecting the small scatter in the collapse times of protohalos with $\delta$ and $R$, or equivalently, the small scatter in $\delta$ of protohalos with $R$ collapsing at $t$ is enough to derive {\it typical} halo properties, though not their scatter.  

\citet{Jea14a} found that, if the density contrast $\delta$ for Gaussian ellipsoidal collapse at $t$, and the r.m.s density contrast $\sigma_0$ of Gaussian peaks of scale $R$ ($\sigma_j$ stands for the $j$-th spectral moment, in general) are written as proportional to their top-hat spherical counterparts, denoted with index `th',\footnote{Factors $D(\ti)/D(t)$ in the top-hat counterparts guarantee that the result is independent of the initial time $\ti$.}, 
\beq
\delta(t,\ti)=r_\delta(t)\delta\F(t,\ti)= r_\delta(t)\,\delta\cc\F(t)\frac{D(\ti)}{D(t)}
\label{deltat}
\eeq
\vspace{-15pt}
\beq
\sigma_0[R(M,t,\ti),\ti]=...=r_\sigma(M,t)\,\sigma_0\F(M,t)\frac{D(\ti)}{D(t)},
\label{rm0}
\eeq
then the proportionality functions $r_\delta$ and $r_\sigma$ are well fitted in all cases of interest by the analytic functions,
\beqa 
r_\delta(t)\approx \frac{a^d(t)}{D(t)}
~~~~~~~~~~~~~~~~~~~~~~~~~~~~~~~~~~~~~~\label{rdelta}\\
r_\sigma(M,t)\approx 1+r_\delta(t){\cal S}(t)\nu\F(M,t)~~~~~~~~~~~~\label{nu}\\
{\cal S}(t)=s_0\!+\!s_1a(t)\!+\!\log\left[\frac{a^{s_2}(t)}{1\!+\!a(t)/A}\right],\,~~~~~~~\nonumber
\label{rs}
\eeqa
for appropriate values of coefficients $d$, $s_0$, $s_1$, $s_2$ and $A$, dependent on cosmology and the halo mass definition adopted. In Equation (\ref{deltat})-(\ref{nu}), $\delta\cc\F(t)$ is the critical linearly extrapolated density contrast for top-hat spherical collapse at $t$, and $\nu$ is the constant peak height, $\delta/\sigma_0$. For simplicity in the notation, we skip from now on $\ti$ in all quantities referring to that arbitrary initial time.

Throughout this Paper we adopt the {\it Planck14} cosmology \citep{P14}, defined by the parameters $(\Omega_\Lambda,\Omega_{\rm m},h,n_{\rm s},\sigma_8,\Omega_b)\!=\! (0.68,0.32,0.67,0.96,0.83,0.049)$, and virial masses, defined by an inner mean density equal to the virial overdensity $\Delta\vir(t)$ \citep{bn98} times the mean cosmic density $\bar\rho(t)$, in which case the coefficients $(d,10^{2}s_0,10^{2}s_1,10^{2}s_2,A)$ take the values $(0.93,2.26,6.10,1.56,11.7)$ \citep{SM24}. 

We remark that this halo-peak one-to-one correspondence does not depend on the detailed mass distribution inside the protohalo, or equivalently, on the aggregation history of the halo. It is the same regardless of whether protohalos are smooth (halos form by pure accretion) or lumpy (they undergo major mergers), or whether protohalos are homogeneous (halos collapse at once) or show some radial slope (they collapse gradually). 

\citet{MSS95} showed that the general relation
\begin{equation}
\frac{\partial\delta({\bf r},R)}{\partial R}=R\nabla^{2}\delta({\bf r},R)\equiv-x({\bf r},R)
\sigma_{2}(R)R,
\label{Gau}
\end{equation}
satisfied by Gaussian smoothing at any point {\bf r} allows one to identify peaks tracing the same accreting halo when the scale $R$ varies according to the mass $M$ of the halo. This way, accreting halos trace continuous peak trajectories in the $\delta$-$R$ plane at $\ti$. According to Equation (\ref{Gau}), the mean peak trajectory $\delta(R)$ of accreting halos with $M\cc$ at $t\cc$ is the solution, for the suited boundary condition, of the differential equation 
\beq
\frac{\der \delta}{\der R}=-\lav x\rav [R,\delta(R)]\,\sigma_2(R)R,
\label{eqtraj0}
\eeq
where 
\beq
\lav x\rav (R,\delta)=\frac{G_1(\gamma,\gamma\nu)}{G_0(\gamma,\gamma\nu)}
\label{hat}
\eeq
is the mean curvature of peaks with $\delta$ at $R$,
\beq
G_i(\gamma,\gamma\nu)\!\equiv \!\int_0^\infty\! \der x\,x^i\,F(x) \frac{{\rm e}^{-\frac{(x-\gamma\nu)^2}{2(1\!-\!\gamma^2)}}}{[2\pi(1-\gamma^2)]^{1/2}},
\label{G}
\eeq
with the function $F(x)$ calculated by \citet{BBKS} (see Paper I), and $\gamma\equiv \sigma^{2}_{1}/(\sigma_{0}\sigma_{2})$.

Given the small scatter in the curvature of peaks, the function $M(t)$ obtained by plugging $\delta(t)$ given by Equation (\ref{deltat}) into $R(\delta)$, inverse of the solution of Equation (\ref{eqtraj0}), and the resulting $R(t)$ into $M(R,t)$ given by the implicit Equation (\ref{rm0}), is close to the mass growth of all accreting halos with $M\cc$ at $t\cc$. Then, the inside-out growth of accreting halos \citep{Sea12a} implies that their mass $M(r)$ inside radius $r$ is the mass $M(t)$ of the progenitor halo at the time $t$ where its virial radius was $r$. Thus, we can determine $M(r)$ from the two latter functions, and differentiating it, we are led to their typical spherically averaged density profile $\rho(r)$. Notice that this derivation of $\rho(r)$, simpler than the one followed in \citet{Sea12a} to prove the inside-out growth of accreting halos, is the same as followed in GSMH, except for the fact that these authors use the mass growth rate $\der M/\der t$ predicted by the EPS model \citep{Rea01}.  

\section{AM Growth of Accreting Halos}
\label{linear}

The AM growth of accreting halos relies on that of protohalos evolving into them, so we need to distinguish between the AM of both kinds of objects. To do this the AM of protohalos is hereafter denoted as $J\ph$ (plus an upper index `L' in case of the Lagrangian value), while the AM of the (progenitor) halos is simply denoted as $J$, as done with their mass $M$ in Section \ref{CUSP}, except for the AM of the final halo with mass $M\cc$ at the collapse time $t\cc$, which is consistently denoted as $J\cc$. In addition, we will make use of the AM of that part of protohalos that is linear at all scales, denoted as $J\p^{\rm lin}$, and of homogeneous protohalos as assumed in Paper I, denoted as $J\p^{\rm h}$.   

As shown by \citet{W84}, the AM of linear non-spherical protohalos in TTT grows by keeping its orientation fixed and the $i$-th Cartesian component equal, at leading order, to 
\beq
(J\ph)_i(t)= - a^2(t)\dot D(t)(J\ph\Lag)_i, 
\label{Jp}
\eeq
where 
\beq
(J\ph\Lag)_i = \epsilon_{ijk}{T}_{jl}{I}_{lk}
\label{1}
\eeq
is its constant Lagrangian counterpart. In Equation (\ref{1}) $\epsilon_{ijk}$ is the fully antisymmetric Levi-Civita rank-three tensor, ${T}_{ij}\equiv  \partial^2 \phi/\partial x_i\partial x_j$ is the tidal tensor at the center of mass (c.o.m.) of the protohalo, and ${\bf I}$ is its inertia tensor with respect to that point. 

In Paper I, Equation (1) was implemented to linear protohalos with $\delta(t\cc)$ and $R(M\cc)$ at $\ti$, subject to the tidal torque of neighboring mass fluctuations, {\it taking into account that they are homogeneous at leading order in the perturbed density}. Averaging their Lagrangian AM for the joint PDF of its arguments, we obtained the mean $J\p\Lag$, and, taking the Lagrangian AM at the most probable values of its arguments, we obtained the median $J\p\Lag$.\footnote{$J\p\Lag$ is lognormal (see Section \ref{AM}), so the logarithm of the median is the most probable value of $\ln J\p\Lag$.} The result was that, to leading order in the mean ellipticity and prolateness, both values essentially coincided, being given by 
\beq
J\ph\Lag= 0.23\frac{G\bar\rho^{1/3}_0 M\ff^{5/3}}{H_0^2 \Omega_0}g\left(\frac{r_{\rm one}}{R\F}\right)^{\!m} \!\frac{\delta(t\cc,\ti)}{D(\ti)},
\label{def}
\eeq
where $G$ is the gravitational constant, $\bar \rho_0$, $H_0$, and $\Omega_0$ are the present mean cosmic density, Hubble constant, and matter density parameter, respectively, $m\equiv -(n+3)/2$, with $n$ being the real or effective power-law index of the power spectrum at $M\cc$, and factors $g$ and $r_{\rm one}/R\F$, the latter giving the separation between the c.o.m. of the protohalo and the main torque source scaled to the protohalo top-hat scale, are the following constant weakly $M\cc$-dependent functions,
\beqa
g= \frac{1}{\left(\lav x\rav^2+\frac{6}{5}\right)^{1/2}}\left[1-\frac{1.182}{\left(\lav x\rav^2+\frac{6}{5}\right)^2}\right]~~~~~~~
\label{g}\\
\frac{r_{\rm one}}{R\F}=\left[8+\frac{3\pi r_{\rm R}^3(n,M\cc,t\cc)}{2\!\left(\!\frac{n\!+\!5}{6}\!\right)^{3/2}\!G_0(\gamma,\gamma\nu){\rm e}^{-\frac{\nu^2}{2}}}\right]^{1/3},~~~
\label{im2}
\eeqa
with $r_{\rm R}^m(n,M\cc,t\cc)$ defined as $\left[K_n\, r_\sigma(M\cc,t\cc)\right]$, where $K^2_n\equiv\int_0^\infty \der x \, x^{n+2}\,(W\F)^2(x)$ $/\int_0^\infty \der x \, x^{n+2}\,W^2(x)$, being $W(x)$ and $W\F(x)$ the Fourier transforms of the Gaussian and top-hat filters. 

Equation (\ref{def}) includes an extra factor $4\pi$ missing in Equation (C6) of Paper I. This correction does not alter, however, the conclusion drawn in that Paper that our predictions agreed with the results of simulations. Indeed, the masses $M$ of simulated halos provided by \citet{Sea00} used to compare our results with were multiplied by a factor 4.5 so as to convert them into virial masses, without touching their AM, $J$. However, as we will see below, $J$ is proportional to $M^{5/3}$, so we should also have multiplied $J$ by a factor $4.5^{5/3}=12.3$, very close to the factor $4\pi=12.6$ multiplying the predicted AM. 

But Equation (\ref{def}) holds for {\it linear} protohalos, while at late times protohalos become non-linear, contract, and undergo shell-crossing. Moreover, protohalos are homogeneous only at leading order in $\delta$. Strictly speaking, their density profile is outward decreasing from the peak, so protohalos become non-linear and collapse gradually from inside out, and, at any time $t$, they harbor a central highly non-linear relaxed core, the progenitor at $t$ of the halo with $M\cc$ at $t\cc$, which progressively accretes the rest of the protohalo. In addition, protohalos usually harbor other massive enough small-scale regions with higher density contrast than average (lumps), which also collapse first, giving rise to multiple progenitors that merge before $t\cc$. We must thus account for all these non-linear effects. 

\subsection{Gradual Monolithic Collapse}\label{PAM2}

The fact that protohalos collapsing monolithically (i.e., with no significant lumps) develop a central relaxed core does not affect their collapse time, dependent as mentioned on their global density contrast only. However, it makes a big difference for their inertia tensor, causing Equation (\ref{def}) to hold only at early times, when the progenitor halo is tiny. At late times, the outer part of the protohalo stays linear at all scales (hereafter simply ``fully linear''), and hence, strictly homogeneous and sensitive to the tidal torque of neighboring mass fluctuations (the central progenitor causes no torque on it), as assumed in Equation (\ref{def}). But, when its innermost shells become non-linear and markedly contract, their inertia tensor rapidly increases, causing their AM to stop growing when they are accreted by the central progenitor halo. As a consequence, the AM of the latter only grows by adding up the aligned frozen AM of accreted shells. 

Therefore, to calculate the AM growth of protohalos collapsing monolithically it is convenient to split them in two parts: i) the central progenitor halo with mass $M(t)$ inferred in Section \ref{CUSP}, whose AM $J(t)$ grows by adding up the frozen AM of accreted shells, and ii) the surrounding fully linear (and homogeneous) part with mass $M\ff-M(t)$, whose AM, $J^{\rm lin}\ph(t)$, grows as calculated in Paper I. 

As shown in Paper I, the proportionality of the Lagrangian AM on mass to ${5/3}$ in homogeneous protohalos arises from their inertia tensor. As the outer fully linear part of the protohalo has a central hole of mass $M(t)$, with essentially the same triaxial shape, orientation, and c.o.m. as the entire protohalo,\footnote{The shape, orientation, and position of real peaks may slightly vary over their trajectories. But, in the absence of lumps, the accretion of shells onto the halo progenitor proceeds in a very symmetric way, so there can be essentially no shift in those properties towards any particular direction.} its AM, $J\ph^{\rm lin}(t)$, is proportional to $M\ff^{5/3}-M^{5/3}(t)$ instead of simply to $M\ff^{5/3}$. Thus, neglecting as usual the virialization time of collapsing shells,\footnote{In structure formation models, the collapse time $t\cc$ of protohalos with $M\cc$ is identified to the time the fully {\it virialized} halo appears.} we are led to
\beq
J^{\rm lin}\ph(t)=J\p^{\rm h}(t) \left\{1\!-\!\left[\frac{M(t)}{M\ff}\right]^{5/3}\!\right\},
\label{Edef2}
\eeq
where $J\p^{\rm h}(t)$ is the AM of the protohalo were it homogeneous (i.e. with Lagrangian value given by Equation (\ref{def}),  as assumed in Paper I. Note that, since $M(t)$ is very similar for the progenitors of all halos with $M\cc$ at $t\cc$, $J^{\rm lin}\ph(t)$ given by Equation (\ref{Edef2}), with the mean (median) value of $J\p^{\rm h}(t)$, is very nearly the {\it mean (median)} AM of their outer fully linear part. 

\begin{figure}
\includegraphics[scale=1.25,bb= 45 0 50 145]{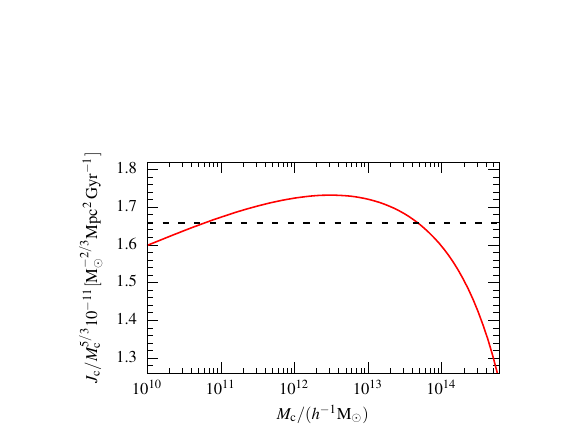}
 \caption{Predicted $J\cc/ M\cc^{5/3}$ relation as a function of halo virial mass (red solid line). The horizontal black dashed line marks the middle value between the minimum and maximum values in the range from $10^{10}h^{-1}$\modot to $10^{14}h^{-1}$\modotc.} 
\label{f0}
\end{figure}

\begin{figure}
\includegraphics[scale=1.25,bb= 43 0 50 200]{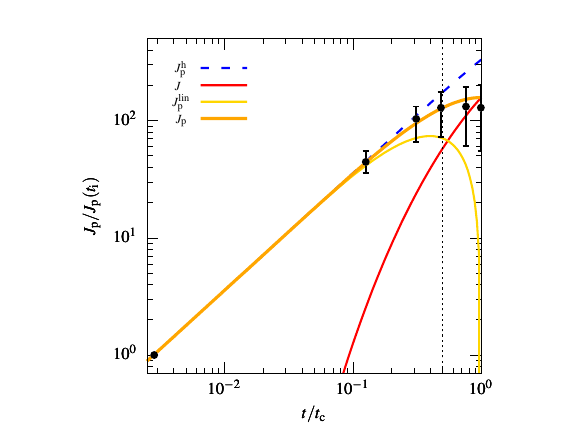}
 \caption{Predicted mean AM growth $J\ph=J+J\p^{\rm lin}$ scaled to the initial value $J\p(\ti)$ at $z\ii=50$ of protohalos collapsing into current halos with virial mass $10^{13}$\modot in the {\it Planck14} cosmology (thick orange line) compared to the mean and 1-$\sigma$ values of simulated protohalos obtained by \citet{Pea02a} (black dots and error bars). Also plotted are the predicted contributions to it from the central progenitor halo, $J/J\p(t_i)$ (red line), and its surrounding fully linear (and homogeneous) part, $J\p^{\rm lin}/J\p(t_i)$ (yellow line). For comparison, we plot the linear evolution until $t\ff$ of the mean scaled AM, $J\p^{\rm h}/J\p(t_i)$, of globally homogeneous protohalos (blue line). The vertical dotted line marks the time $t/t\cc=0.5$.}
\label{f1}
\end{figure}

Over that outer part, the specific AM is kept aligned with uniform value of $J\ph^{\rm lin}(t)/[M\ff^{5/3}-M^{5/3}(t)]$ because all the centered ellipsoids are homothetic, and the tidal torque they suffer is due to the same external source acting on the entire protohalo, i.e.,  at the same fixed separation $r_{\rm one}/R\F$ from their c.o.m.\footnote{There can be no internal torque source because this would mean that the protohalo is lumpy, contrarily to what is assumed.} Consequently, the AM of the progenitor halo, which grows by simply adding up the frozen AM of accreting shells, satisfies the differential equation
\beq
\frac{\der J}{\der t}\!=\! \frac{J\ph^{\rm lin}(t)}{M\ff^{5/3}\!-\!M^{5/3}(t)} \frac{\der M^{5/3}}{\der t}
\!=\!J\p^{\rm h}(t) \frac{\der\! \left(\!\frac{M}{M\cc}\!\right)^{\!\!5/3}}{\der t},
\label{eq}
\eeq
whose solution for the initial condition $J(0)=0$ is 
\beqa
J(t)= Q(t) J\ph\Lag~~~~~~~~~~~~~~~~~~~~~~~~~~~~~~~~~~~~~~~~~\label{Jt}\\
Q(t)=\int_0^t \der \tilde t\,a^2(\tilde t)\dot D(\tilde t)\frac{\der\left(\!\frac{M}{M\cc}\!\right)^{\!\!5/3}}{\der \tilde t}.~~~~~~~~~~~~~~~~\label{Q}
\eeqa
The small variations in $\der (M/M\cc)^{5/3}/\der t$ of individual halos is smoothed out by the time integral, so $Q(t)$ is again very similar for all accreting halos with $M\cc$ at $t\cc$, and $J(t)$ given by Equation (\ref{Jt}), with $J\p\Lag$ equal to the mean (and median) Lagrangian AM of protohalos at early times (Equation (\ref{def})), gives very nearly their {\it mean} (and {\it median}) progenitor AM growth. 

The function $Q(t)$ appears to be little dependent on $M\cc$, so $J(t)$ is nearly proportional to $M\cc^{5/3}$ through $J\p\Lag$. That is the case, in particular, of the final halo AM, $J\cc\equiv J(t\cc)$, which agrees with the results of simulations \citep{EJ79,BE87,Cea96b,Sea00,Liao17}. Concretely, as shown in Figure \ref{f0}, $J\cc/M\cc^{5/3}$ is constant to better than $\sim 3.5$\% over 4 orders of magnitude (from $10^{10}h^{-1}$\modot to $10^{14}h^{-1}$\modotc). At the high mass end it shows a substantial decrease, but at those masses the halo abundance falls off exponentially. In fact, it is this fall of the halo abundance at high masses what causes a marked increase of $r_{\rm one}/R\F$ there (see Paper I), and the consequent rapid decrease of $J\cc/M\cc^{5/3}$. 

We remark that, as far as $J\p\Lag$ was calculated in Paper I to leading order in several quantities, $J\cc$ is approximate. However, all quantities scaled to $J\cc$, such as those mentioned in Section \ref{specific}, or to $J(\ti)$ for any arbitrary cosmic time $\ti$ (see next) should be accurate. Moreover, since the dependence of $J(t)$ on $M\cc$ is the same as in $J\cc$, and $Q(t)$ is little dependent on cosmology (for all cosmologies of interest), such scaled quantities should also be nearly universal. 

In Figure \ref{f1}, we see that the mean protohalo AM, $J\p(t)=J^{\rm lin}\ph(t)+J(t)$, coincides at early times with the mean AM of homogeneous objects, $J\p^{\rm h}(t)$, calculated in Paper I. (In this sense, the protohalo Lagrangian AM $J\p\Lag$ is constant, indeed, at those times, and well estimated in Paper I.) However, when a large enough fraction of the protohalo has been accreted onto the central progenitor halo, $J^{\rm lin}\p(t)$ stops growing and rapidly falls off, causing $J\p(t)$ to deviate from $J\p^{\rm h}(t)$ and flatten. The result is a predicted (nearly universal) scaled protohalo growth, $J\p(t)/J\p(\ti)$, that agrees with that found by \citet{Pea02a}\footnote{These authors provide the protohalo AM scaled to $J\cc$ and to $J(t)$ at $z_i=50$. In Figure \ref{f1} we show the latter because the former hides the scatter in the AM of the simulated halos at $t\cc$. Yet, protohalos also have different initial AM values, so part of the scatter is also hidden in the chosen representation} (see also \citealt{Sea00}). Note, in particular, that $J\p^{\rm h}(t)$ reaches the AM $J\cc$ of the final halo between $t\cc/3$ and $t\cc/2$, as found in simulations.

Obviously, since the accreting progenitor halo grows inside-out, with its AM permanently aligned, the final halo must end up with roughly spherical\footnote{Relaxation increases the sphericity of triaxial systems; \citep{Sea12b}.} shells rotating as rigid bodies, with outward increasing specific AM aligned with the total {\bf J}$\cc$ axis as found in simulations \citep{Bea01,Bea07,Bea10}. We will comeback to the consequences of this prediction in Section \ref{specific}.

But, before proceeding, a caveat is in order. Our model relies on the assumption that the orientation and Lagrangian modulus of the protohalo AM are kept fixed (Equation (\ref{Jp})), or, in terms of our treatment in Paper I, that the shape and orientation of the protohalo and the main tidal torque source and their comoving separation stay constant. That is indeed a good approximation in the linear regime. However, when the shells of the fully linear part of the protohalo (and in the main tidal torque source) leave the linear regime, all these quantities slightly change. In particular, their non-null angular momentum causes them to spin more markedly, tending to align their inertia tensor with the tidal tensor. This, in turn, causes not only the direction of their AM to smoothly slip, but its modulus also to decrease relative to the the linear evolution. But these changes can only take place during the small time interval between the shells leaving the linear regime and their AM freezing out, so these effects are expected to be small. This is why, for simplicity, we have neglected them. However, they must be borne in mind as they likely explain why, in simulations, the modulus of the protohalo AM is found to even slightly decrease when $J\p^{\rm lin}(t)$ falls off (see Figure \ref{f1}) and its orientation to very smoothly change since the beginning \citep{Pea02a,Vea02,Wea17,Lea21,Lea25}.

\subsection{Major Mergers}\label{PAM1}

As mentioned, simulations show that those halo properties that directly arise from their collapse and virialization, i.e., affecting in the same way all the components of those systems, do not depend on their aggregation history. The reason for this is that violent relaxation produced in major mergers yields the loss of information on lumps present in protohalos at the base of those mergers so that the most probable configuration reached (i.e., the equilibrium configuration) is identical to the typical configuration arising from monolithic collapse \citep{SM19}. This is why to derive halo properties such as their density profile we have the right to assume monolithic collapse or pure accretion. 

But, as discussed in \citet{SM19} and also confirmed by simulations \citep{WW09,Bea12}, this conclusion does not hold for those properties involving interactions, such as tidal stripping or dynamical friction, that distinguish between halo constituents with different properties (mass, shape, location,...). These properties do keep the memory of major mergers when the spatial distribution of such constituents was rearranged (e.g., \citealt{III,IV}). 

Lumps in protohalos feel not only the tidal torque of external mass fluctuations, but also the tidal torque of other lumps, meaning that the inner rotational properties of {\it protohalos} arise from interactions between constituents (lumps) with different properties. Thus, we might wonder whether the rotational properties of {\it halos} also keep information about major mergers. But this is not the case. When two progenitors of a halo arising from two lumps in the protohalo merge, the part of their AM due to their mutual tidal torque cancels, and the only AM that survives is that due to the external torque source acting on their composite system, as if there have been no lumps. Therefore, violent relaxation yields the memory loss not only of the mass and location of preexisting lumps, but also of their mutual tidal torque, and the typical rotational halo properties can also be inferred assuming monolithic collapse, as done in Section \ref{PAM2}. 

This conclusion is not contradictory with the frequent, often dramatic, change in the AM modulus and orientation of halos found in simulations when monitoring {\it their individual evolution} (e.g., \citealt{Bea10,Bea16,Gea21,EB22}). These changes are caused by major mergers between protohalo progenitors. Indeed, in the case of clumpy protohalos, the AM of the progenitors formed from these clumps is essentially due to the tidal torque exerted by their merging partners which are the closer mass fluctuation of the same mass (Paper I). But the AM of essentially opposite sign of both objects caused by their mutual torque cancels when they merge (actually, since their first fly by). Consequently, after the merger, the AM of the final object is due to their common external torque source. This yields an important change in the modulus and orientation of their individual AM (even when the protohalo including both objects is still in linear regime). However, the global AM of the composite system evolves continuously as if there were no merger. This is why simulations following the evolution of the AM of the protohalo grows according to Equation (\ref{1}) until freezing out, as if there were no major mergers (e.g., \citealt{Sea00,Pea02a}), despite the changes in the AM of the inner individual halos.

\section{Halo Spin}\label{AM}

The dimensionless spin parameter \citep{P69}, 
\beq
\lambda=\frac{J\ff|E\ff|^{1/2}}{GM\ff^{3/2}},
\label{lambdac}
\eeq
measures the importance of rotation in the energetics of halos. Unfortunately, this parameter involves not only the AM of the halo, but also its total energy $E$ (including the rotational component), density profile, and triaxial shape, making it difficult to calculate in simulations. This is why \citet{Bea01} introduced the alternate spin parameter  
\beq
\lambda'=\frac{J\ff}{\sqrt{2}M\ff R\ff V_{\rm cir}},
\label{lambdap}
\eeq
where $R\ff$ is the radius of the halo and $V_{\rm cir}=(GM\ff/R\ff)^{1/2}$ its circular velocity, which for virial masses takes the form
\beq
\lambda'=\frac{1}{\sqrt{2G}}\left[\!\frac{4\pi}{3}\bar\rho(t\ff)\Delta\vir(t\ff)\right]^{1/6}\! \frac{J\ff}{M\ff^{5/3}}.\!
\label{lm}
\eeq
Both spin estimates coincide for spherically symmetric objects endowed with an isothermal density profile, while for halos with the NFW density profile \citep{NFW95} of concentration $c$ we have $\lambda'\approx \lambda h(c)$, where $h(c)=0.5c[(1+c)^2-1-2(1+c) \ln (1+c)]/[c-(1+c) \ln (1+c)]^2\approx [2/3+(c/21.5)]^{0.7}$ \citep{Mo98}. 

Equations (\ref{lm}) and (\ref{Jt}) state that the spin $\lambda'$ of halos with $M\cc$ at $t\cc$ is proportional to the Lagrangian AM, $J\p\Lag$, with a fixed proportionality factor. And, since $J\p\Lag$ is a positive function, product of many independent random quantities (see Paper I), the central limit theorem indicates that $\ln J\p\Lag$ should be close to normally distributed, and hence, $J\p\Lag$ close to lognormally distributed. Thus, it is well understood that $\lambda'$ is also (\citealt{BE87,CL96,VdB02,Aea05,GY07,Mea07,Zs17}; see also \citealt{Bea07}). Moreover, the mean and median spin, $\lambda'_{\rm mean}$ and $\lambda'_{\rm med}$, of halos with $M\cc$ at $t\cc$ are proportional to the mean and median $J\p\Lag$ calculated in Paper I. 

In addition, since $J\cc/M\cc^{5/3}$ is approximately constant in mass and the proportionality factor in Equation (\ref{lm}) is little dependent on time, $\lambda'_{\rm mean}$ and $\lambda'_{\rm med}$ must be approximately universal, as also found in simulations \citep{BE87,CL96,VdB02,Aea05,Bea07,GY07,Mea07}. 

Plugging the ratio $J\cc/M\cc^{5/3}$ derived in Section \ref{PAM2} into Equation (\ref{lm}) leads to a mass-weighted median spin of $\lambda'_{\rm med}= 0.035$ for current halos with virial masses in the range $12\le \log [M\cc/(h^{-1}$\modotc$)]\le 14$ in the {\it Planck14} cosmology. Despite the approximations involved in the calculation of $J\p\Lag$, this value surprisingly matches that, $\lambda'_{\rm med}=0.035\pm 0.005$, found in simulations for the same halo masses and essentially the same CDM cosmology by \citet{Bea01} (see also \citealt{BE87,VdB02,Aea05,Mea07}). Similarly, using the Millennium simulation, \citet{Bea07} found $\lambda'_{\rm med}=0.03687\pm 0.000016$ for halos with virial masses in the range $10\le \log [M\cc/(h^{-1}$\modotc$)]\le 15$ in the {\it WMAP} cosmology. What is more interesting, they also found a slight trend for high-mass halos to have lower spins (see also \citealt{CL96,Mea07}), a trend we also find towards $10^{15}h^{-1}$\modot (see Figure \ref{f0}). 

Certainly, the prediction $\lambda'_{\rm mean}\approx \lambda'_{\rm med}$ also implies\footnote{If $X$ is lognormal with $\mu\equiv \lav \ln X\rav$ and $\sigma^2\equiv\lav (\ln X-\mu)^2\rav$, then the median and mean $X$ values are $X_{\rm med}=\exp(\mu)$ and $\lav X\rav=\exp(\mu+\sigma^2/2)$.} a dispersion in $\ln\lambda'$ of $\sigma_{\ln \lambda'}\la 0.20$, while \citet{Bea01} and \citet{Bea07} found $\sigma_{\ln \lambda'}=0.50\pm 0.03$ and $\sigma_{\ln \lambda'} = 0.5103 \pm 0.00028$, respectively. But this is unsurprising because the mean and median $J\p\Lag$ calculated in Paper I referred to protohalos {\it with the most probable $\delta$ for ellipsoidal collapse at $t\cc$}, not to protohalos {\it collapsing at} $t\cc$, so the dispersion in $\ln J\p\Lag$ did not include the scatter in $\delta$ mentioned in Section \ref{CUSP}.\footnote{The median $J\p\Lag$ derived in Paper I is correct because it should be evaluated at the most probable $\delta$, anyway. However, the mean $J\p\Lag$ is slightly underestimated. Unfortunately, it cannot be better estimated because of the much more complex joint PDF of its arguments (then including the $\delta$ and $x$ values of protohalos and tidal torque sources) and the unknown $\delta$-PDF.} 

\section{Specific AM Profile}\label{specific}

\begin{figure*}
\includegraphics[scale=1.25,bb= 33 00 50 200]{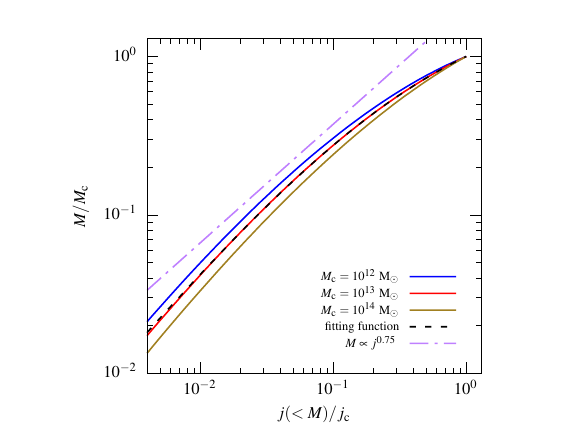}
\includegraphics[scale=1.25,bb= -153 00 50 200]{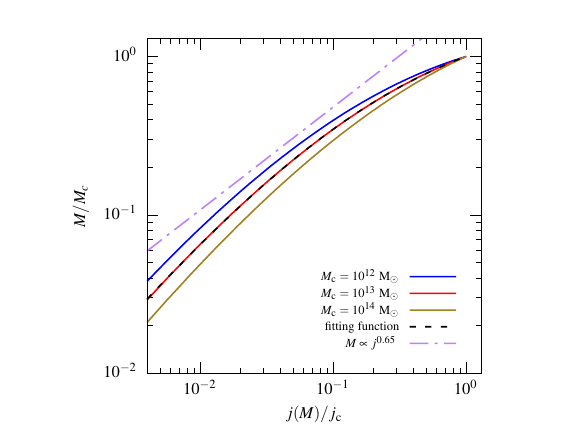}
 \caption{Left panel: Predicted mass of centered spheres as function of their integrated specific AM scaled to its maximum value, $j\cc=J\cc/M\cc$, at the radius $R\cc$ of the halo, in current halos of several virial masses (colored lines), compared to the best fit for $M\cc=10^{13}$\modot to the three-parametric function $\mu M\ff /[1+(j_0/j)^\gamma]^{1/\gamma}$, with $\mu=2.57$, $j_0=0.504$, and $\gamma=0.553$ (black dashed line). Right panel: Same as left panel, but for the local specific AM scaled to its maximum value, $j\cc=j(R\cc)$, at the radius of the halo, compared to the best fit for $M\cc=10^{13}$\modot to the same analytic function, with  $\mu=2.12$, $j_0=0.230$, and $\gamma=0.513$.} 
\label{f2}
\end{figure*}

\begin{figure*}
\includegraphics[scale=1.25,bb= 33 0 50 200]{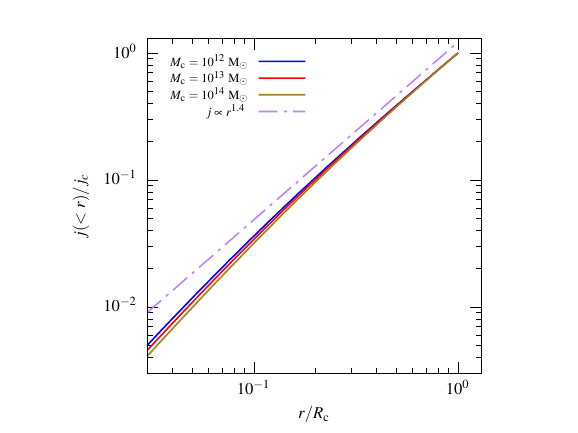}
\includegraphics[scale=1.25,bb= -153 0 50 200]{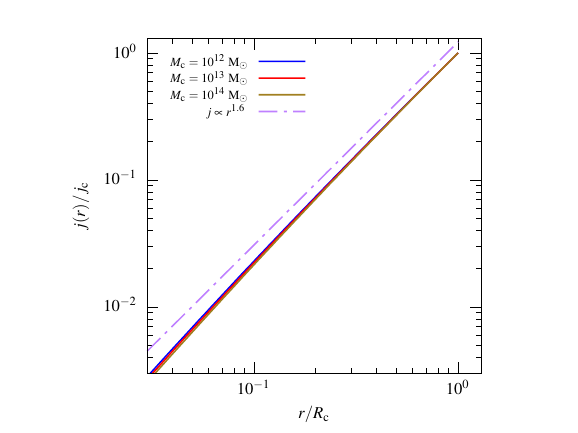}
 \caption{Left panel: Predicted integrated specific AM profile scaled to its maximum value, $j\cc=J\cc/M\cc$, for current halos of several virial masses (solid colored lines). Right panel: Same as left panel, but for the local specific AM profile scaled to the maximum value, $j\cc=j(R\cc)$, at the radius $R\cc$ of he halo.}
\label{f3}
\end{figure*}

The inner specific AM of halos inside radius $r$ or the mass $M\equiv M(r)$ readily follows from their mass and AM growths derived in Sections \ref{CUSP} and \ref{linear}, respectively, and the fact that, in pure accretion, these objects grow inside-out, keeping the AM aligned. 

Indeed, the specific AM of centered spheres of mass $M$, $j(<\!M)$, is simply $J(M)/M$, where $J(M)$ is the AM of the progenitor halo at the time $t$ where its mass $M(t)$ reached $M$. That in turn implies a spherically averaged specific AM profile, $j(r)$, related to the AM of centered spheres of radius $r$ through 
\beq
J(r)=\int_0^{r} \der \tilde r\, 4\pi\, \tilde r^2 j(\tilde r)\rho(\tilde r)
\label{Jr}
\eeq 
with $J(r)$ equal to the AM $J(t)$ of the progenitor at the time its mass $M(t)$ reached $M$. Differentiating Equation (\ref{Jr}) leads to (see Equation (\ref{eq}))
\beq
j(r)=\frac{\der J(r)/\der r}{\der M(r)/\der r}\!=\frac{5}{3}J\p^{\rm h}(r) \left[\frac{M(r)}{M\cc}\right]^{2/3},
\label{jr}
\eeq
with $J\p^{\rm h}(r)$ equal to the AM of the homogenous protohalo at the time $M(t)$ reached $M$, implying in turn the simple relation $j(M)=\der J/\der M=5/3 J\p^{\rm h}(M) [M/M\cc]^{2/3}$ for the local intrinsic AM at the edge of the centered sphere of mass $M$. This derivation of $j(r)$ (specifically, the first equality in Equation (\ref{jr}) is the same as that used by GSMH, although $\der J(r)/\der r$ was inferred there from the AM growth rate implied by a constant halo spin. Since all accreting halos with $M\cc$ at $t\cc$ have very similar mass growths $M(t)$, and $J(t)$ gives their mean AM growth, the previous $j(<\!M)$ and $j(M)$ relations and $j(<r)$ and $j(r)$ profiles are also the {\it mean} relations and profiles of halos with $M\cc$ at $t\cc$. 

In Figure \ref{f2} we plot the predicted cumulative mass-PDF as functions of the integral (left panel) and local (right panel) specific AM, inverse of the predicted mean $j(<\!M)$ and $j(M)$ relations, respectively. \citet{Bea01} found that the $M(j)$ relation of individual simulated halos 
is well fitted to the two-parametric function $\mu M\ff/ (1+j_0/j)$, and this is indeed the case for the $M(j)$ relation (and the $M(<\!j)$ relation as well) derived here, though we find that the three-parametric function $\mu M\ff /[1+(j_0/j)^\gamma]^{1/\gamma}$, with $\gamma\approx 0.5$, with identical asymptotic behavior at both ends, yields an even better fit. Unfortunately, the small fluctuations along the $M(j)$ relation of individual simulated halos (see Figure 4 of \citealt{Bea01}) prevent us from assessing whether this new analytic function also better fits those individual relations or that only happens for the predicted {\it mean} AM, as would be the case if the parameters $\mu$ and $j_0$ of individual objects are correlated enough.\footnote{The scaled relations $M(j\lo/j_0)/M_0$, with $M_0=\mu M\cc/2$, of individual simulated halos is found to be universal \citep{Bea01}, so its inverse $j(M/M_0)/j_0$ is also and coincides with its mean. Thus, {\it provided only} the correlation between $j(M\cc/M_0)$ (equal to $j(2/\mu)$, so a function of $\mu$ only) and $j_0$ is small, will the mean $j(M/M_0)/j_0$ be close to the mean $j(M/M_0)$ times a certain value $1/j_0$ equal to the mean $1/j_0$ of halos, and its inverse $M(j\lo)$ admit the same fit to the analytic function $\mu M\cc /(1+j_0/j)$ as the $M(j)$ relation of individual halos.} 

The predicted mean $j(<r)$ and $j(r)$ profiles are shown in Figure \ref{f3}. There is in the literature no precise profiles of this kind drawn from simulations to compare with. However, as far as the mean $j(<\!M)$ and $j(M)$ relations are correctly predicted and the mass profile $M(r)$ is also, they should too. 

Numerical studies only provide power-law approximations to such profiles and the AM-mass relations. Under this approximation, $\der J/\der M$ is proportional to $J/M$, which implies $j(<\!\!M)\propto j(M)\propto M^\beta$ and, provided $M(r)$ is also approximated by a power-law, $j(<r)\propto j(r)\propto r^\alpha$. In Figures \ref{f2} and \ref{f3}, we see that the power-law approximation is quite rough for $M(j)$ and $M(<\!j)$, but the shape of $M(r)$ conspires with that of $j(<\!M)$ and $j(M)$ to render the $j(<r)$ and $j(r)$ profiles much closer to power-laws. The values $\beta\approx 1/0.75=1.3$ and $\alpha\approx 1.4$ we find for the predicted mean global AM, and $\beta\approx 1/0.65=1.5$ and $\alpha\approx 1.6$ for the predicted mean local AM\footnote{The different value of the indexes found for the global and local AM is due, of course, to the fact that the power-law approximation is not very accurate.} are consistent with the {\it mean} values $\beta=1.3\pm 0.3$ and $\alpha=1.1\pm 0.3$ found in simulations (\citealt{Bea01}; see also \citealt{BE87,Bea10}). Certainly, the former tend to be slightly larger than the latter, but that is unsurprising because the mean index of power-law-approximated functions is necessarily smaller than the index of the power-law-approximated mean function (see the Appendix). In this sense, the predicted $\beta$ and $\alpha$ values would be more representative of the typical values of such indexes in real halos than the mean $\beta$ and $\alpha$ values reported in the literature. 

Lastly, as halos are formed by roughly spherical shells rotating as rigid bodies around the {\bf J} axis (Section \ref{PAM2}), the specific AM at the point {\bf r}, $j\lo({\bf r})$, takes the form
\beq
j(r,\theta)= r^2\sin^2(\theta)\,\omega(r),
\eeq
where $\theta$ is the zenithal angle with the $z$ axis along ${\bf J}$, and $\omega(r)$ is the angular velocity of the shell at $r$. That is indeed the form of $j({\bf r})$ found in simulations \citep{Liao17}. Integrating this specific AM over the azimuthal and zenithal angles, $\phi$ and $\theta$, we find
\beq
\omega(r)=\frac{j(r)}{\pi^2 r^2}.
\label{omega}
\eeq
The mean value $\alpha\approx 1.1$ found in simulations implies $\omega(r)\propto r^{-0.9}$ \citep{Bea10}. However, according to the preceding discussion, the profile $\omega(r)\propto r^{-0.4}$ arising from the predicted (actually quite accurate) value $\alpha\approx 1.6$ would be more representative of the typical angular velocity profile of halos.

\section{Summary and Conclusions}
\label{sum}

Halos grow by alternating periods of smooth accretion and major mergers. As pointed out by \citet{ssm} and \citet{Rea98}, the configuration of equilibrium (i.e., the most probable configuration) of halos set by violent relaxation after a major merger coincides with the mean configuration of halos of the same mass and time grown by monolithic collapse or pure accretion \citep{SM19}. Therefore, the typical properties of these objects arising from their collapse and virialization do not depend on their aggregation history, and to derive them we have the right to assume the simplest case of pure accretion, where halos virialize orderly from inside out and develop this way \citep{Sea12a}. We have shown that this conclusion holds not only for the structural and kinematic properties, but also for the rotational ones.

GSMH showed how to take advantage of that right to infer all those halo properties from their mass and AM growth rates, using the EPS model and assuming a constant spin. Some years later, we followed the same approach in the peak model, which allowed us to reproduce and explain the structural and kinematic halo properties \citep{Sea12a,Sea12b} found in simulations. But to derive their rotational properties we needed first to implement TTT in the peak model, developed precisely to this end. This was achieved in Paper I, where we calculated the AM growth of protohalos that stay linear at all relevant scales, and hence, homogeneous to leading order in perturbed quantities. Here, we have extended this result to protohalos that progressively leave the linear regime, contract, and virialize. 

To do that, we have taken into account that, even in the full linear regime, the density profile of protohalos is slightly decreasing outward from the location of the peak. This causes them to collapse and virialize (through ordered shell-crossing) gradually from the inside out. As a consequence, they harbor at any time a previously collapsed and relaxed core, which is nothing but the progenitor of the final halo, which progressively grows by accreting the rest of the protohalo. Since that outer part is linear at all scales, its AM grows by the effect of the tidal torque of the surrounding mass fluctuations exactly as described in Paper I, except for the fact that it has a central triaxial hole, previously occupied by collapsed shells. Instead, the central progenitor halo has contracted so much that it is no more sensitive to that tidal torque, so its AM grows by just adding up the frozen AM of the accreted shells. 

The linked AM growth of both parts of the system has been monitored until the full collapse of the halo, neglecting the small change in the orientation and shape of protohalo shells (or the main tidal torque source) in the small time interval since they leave the linear regime until their AM freezes out. This way we have been led to an AM growth of the global system that fully reproduces that found in simulations. We have shown that the AM of early protohalos, calculated in Paper I, gives rise to a nearly universal lognormal distribution of halo spins, with a mean consistent with that found in simulations, even in the slight trend for very massive halos to have lower spins. Lastly, we have shown that the inside-out growth of the accreting progenitor halos explains that the structure of simulated halos can be basically described as a series of embedded concentric shells rotating as rigid bodies around the same fixed AM axis. As explained, such a structure of {\it final halos} (with their inner specific AM essentially aligned everywhere) found in simulations is not contradictory with the dramatic changes in the AM modulus and orientation of halos also found in simulations when monitoring their individual evolution. Such changes do take place when halos merge with other halos inside their common linear protohalo, causing the part of their previous AM due to their mutual tidal torque to be neutralized. Subsequently, the AM of the composite system, which is due to the common external tidal torque, keeps on evolving continuously as predicted in pure accretion.

The predicted final structure of halos allows one to derive with unprecedented accuracy the mean (spherically averaged) specific AM distribution in mass and radius, which are also fully consistent with those found in numerical studies (often approximated by simple the power-laws). All these results demonstrate that, as guessed, TTT fully accounts for the rotational properties of simulated dark matter halos. 

With the present study we culminate the work done over the last decade showing that all CDM halo properties found in cosmological $N$-body simulations (including their substructure, mass function and primary and secondary biases) can be reproduced analytically, often in more detail, and explained in the peak model of structure formation. The fact that all the properties of simulated halos arise from the initial perturbed density field (determined by its power spectrum) is, of course, unsurprising as it is the essence of simulations. But the fact that they are successfully predicted, with no free parameter, from peak statistics was not obvious. It demonstrates that peaks are robust halo seeds, the inside-out growth of accreting halos and the ignorable effects of major mergers making the rest. 

\begin{acknowledgments}
This work was funded by the Spanish MCIN/AEI/ 10.13039/501100011033 through grants CEX2019-000918-M (Unidad de Excelencia `Mar\'ia de Maeztu', ICCUB) and PID2022-140871NB-C22 (co-funded by FEDER funds) and by the Catalan DEC through the grant 2021SGR00679.
\end{acknowledgments}



{}

\appendix

\section{Typical Index of Power-law-Approximated Functions}

The power-law approximation $f\ii(x)\approx A\ii x^{\alpha\ii}$ of $N$ functions $f\ii(x)$ leads to the relation
\beq
\alpha\ii\approx \frac{1}{\ln x}\ln\frac{f\ii(x)}{A\ii}.
\label{first}
\eeq
Provided all these functions $f\ii(x)$ are close enough to their mean, $\lav f\ii(x)\rav$, so that they can be expressed as the mean times one plus a residual $\epsilon\ii(x)\ll 1$, the mean of indexes $\alpha\ii$ takes the form
\beq
\lav \alpha\ii\rav\approx \frac{1}{\ln x}\frac{1}{N}\sum_{i=1}^N \ln\frac{f\ii(x)}{A\ii}=\frac{1}{\ln x}\frac{1}{N}\sum_{i=1}^N \ln\frac{\lav f\ii(x)\rav \left[1+\epsilon\ii(x)\right]}{A\ii}.
\eeq
And, Taylor expanding $\ln[1+\epsilon\ii(x)]$ to second order, we arrive at
\beq
\lav \alpha\ii\rav\approx \frac{1}{\ln x}\left[\frac{\ln \lav f\ii(x)\rav}{A}-\frac{1}{2N}\sum_{i=1}^N \epsilon^2\ii(x)\right]\approx \alpha-\frac{1}{2}\frac{\lav \epsilon^2\ii(x)\rav}{\ln x},
\label{alpha}
\eeq
where we have taken into account the equality $\lav\epsilon\ii(x)\rav=0$ and adopted the power-law approximation of the mean $f\ii$, $\lav f\ii(x)\rav(x)\approx Ax^{\alpha}$, with $\ln A=\lav \ln A\ii\rav$ (neglecting the residuals should lead to $\lav \alpha\ii\rav\approx \alpha$). Note that Equation (\ref{alpha}) is approximated not only because of the truncated Taylor expansion, but also because of the neglect of the small $x$-dependence of the right-hand member, like in the definitions of $\alpha\ii$ (Equation (\ref{first})) and $\alpha$. 

Equation (\ref{alpha}) states that $\lav \alpha\ii\rav$ equals $\alpha$ to first order in the residuals, but it is smaller than $\alpha$ to any higher order. Consequently, if $f\ii(x)$ are close to their mean $\lav f\ii\rav(x)$, as assumed, $\alpha$ is more representative of their typical power index than $\lav \alpha\ii\rav$, which underestimates it. 



\end{document}